# Imaging of Spatially Extended Hot Spots with Coded Apertures for Intra-operative Nuclear Medicine Applications

**I. Kaissas**[a,b]**, C. Papadimitropoulos**[c]**, C. Potiriadis**[b]**, K. Karafasoulis**[d]**, D. Loukas**[e] **and C. P. Lambropoulos** [c,1]

[a] *Aristotle University of Thessaloniki,*
*Egnatia Str., University Campus, Dept. of Electrical & Computer Engineering, 54124, Thessaloniki, Greece*

[b] *Greek Atomic Energy Commission,*
*Patriarxou Grigoriou & Neapoleos, Agia Paraskevi- Attiki, 15310 Greece*

[c] *Technological Educational Institute of Sterea Ellada ,*
*Psahna-Evia, 34400 Greece*

[d] *Hellenic Army Academy,*
*Vari, Attiki, 16673 Greece*

[e] *Institute of Nuclear Physics, National Center for Scientific Research Demokritos,*
*Patriarxou Grigoriou & Neapoleos, Agia Paraskevi- Attiki, 15310 Greece*
   *E-mail*: `lambrop@mail.teiste.gr`

ABSTRACT: Coded aperture imaging transcends planar imaging with conventional collimators in efficiency and Field of View (FoV). We present experimental results for the detection of 141keV and 122keV $\gamma$-photons emitted by uniformly extended $^{99m}$Tc and $^{57}$Co hot-spots along with simulations of uniformly and normally extended $^{99m}$Tc hot-spots. These results prove that the method can be used for intra-operative imaging of radio-traced sentinel nodes and thyroid remnants. The study is performed using a setup of two gamma cameras, each consisting of a coded-aperture (or mask) of Modified Uniformly Redundant Array (MURA) of rank 19 positioned on top of a CdTe detector. The detector pixel pitch is 350 μm and its active area is $4.4\times4.4$ cm$^2$, while the mask element size is 1.7mm. The detectable photon energy ranges from 15 keV up to 200 keV with an energy resolution of 3-4 keV FWHM. Triangulation is exploited to estimate the 3D spatial coordinates of the radioactive spots within the system FoV. Two extended sources, with uniform distributed activity (11 and 24 mm in diameter, respectively), positioned at 16cm from the system and with 3cm distance between their centers, can be resolved and localized with accuracy better than 5%. The results indicate that the estimated positions of spatially extended sources lay within their volume size and that neighboring sources, even with a low level of radioactivity, such as 30 MBq, can be clearly distinguished with counting time about 3 seconds.

---

[1]Corresponding author.

# Contents

## 1. Introduction

Several surgeons transact intra-operative procedures for localizing $^{99m}$Tc-traced cancerous tissues assisted by gamma counter probes or small mobile gamma cameras. The actual size of sentinel nodes and thyroid remnants traced with radio-isotope is from a few millimeters up to 2 cm in diameter [1, 2]. Considering, also, the anatomy of these tissues, they are spaced apart more than 2 cm. These spatial characteristics are convenient for coded aperture imaging.

In this study we investigate the performance of a system, described in detail elsewhere [3, 4], to localize and spatially distinguish $^{99m}$Tc extended sources, by taking into account their spatial extent. Two different mask pitches were designed to operate optimally at specific distances from the system. In the following sections, a description of the system along with the experimental and simulated results are shown for each case.

## 2. Experimental Setup, Methods and Results

The setup consists of two coded apertures mounted on top of two energy dispersive position sensitive detectors (PSD) developed by AJAT Oy [5]. Each PSD contains an array of 4×2 hybrids with a total active area 4.4×4.4 cm$^2$. One hybrid consists of a CdTe pixilated crystal, 1mm thick, bump bonded on a CMOS ASIC with 350μm pixel pitch. The PSD records the charge delivered to pixels by each photo-conversion and sends it for online processing to a PC at a rate of ~27frames/s. Thus, when a radioactive source irradiates the mask-PSD system, a fraction of the mask is projected onto the PSD's surface forming a 2D image (shadowgram). Subsequently, the correlation matrix is defined as the cross-correlation of the shadowgram with a predefined, mask-like, decoding function [6, 7]. The direction of the source is revealed as a peak on the correlation matrix. Combining the directions provided by the two mask-PSD systems, triangulation yields the 3D coordinates of the source. The origin of the system reference frame (SRF) is in the middle between the PSDs (Figure 1, left). The distance between the PSDs is adjustable.

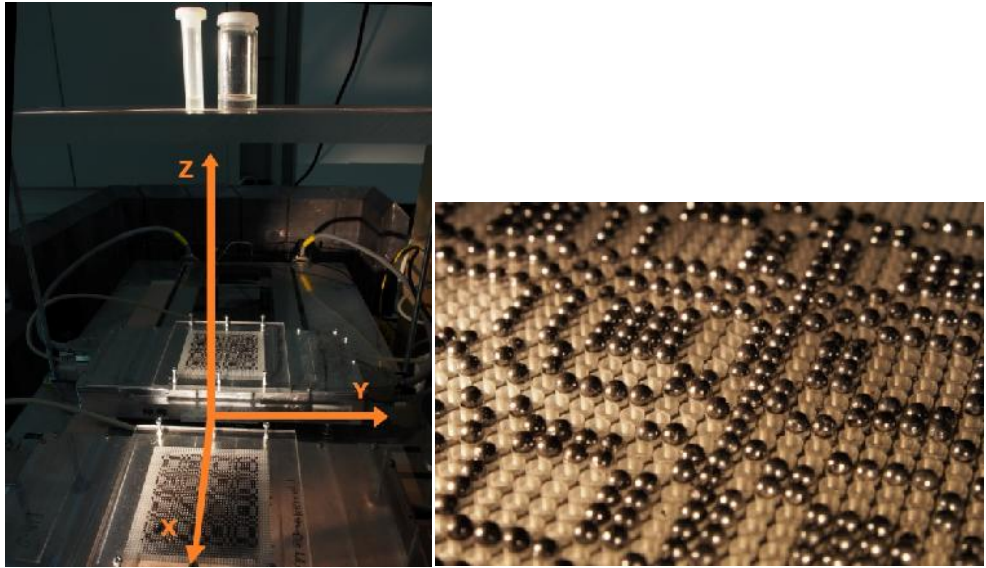

**Figure 1.** An actual photograph of the two coded aperture gamma cameras and the two cylindrical containers (Type A and Type B) positioned on top of a height-adjustable seat of plexiglas and filled up to 9 mm of $^{99m}$Tc solution (left). A close-up detail of MURA mask 19R-1821 presenting the mask elements made of Pb spheres (right).

The $Z$ coordinate of the source position along with the mask-to-detector distance $b$ define the magnification factor $m$ ($m = 1 + \frac{b}{z-b}$) that affects the dimensions of the projection of every mask element on the detector surface and consequently the number of mask elements on the shadowgram. Shadowgrams with number of elements different than the rank of the mask (i.e. 19x19) result in undesired side-lobes, shown in Figure 2c beside the main peak on the correlation matrix [6, 7]. Keeping $b$ constant at 20 mm, two pairs of MURA masks of rank 19 were developed and utilized. One mask has 1.821 mm element pitch in order to minimize the side-lobes ideally for a source located at 160 mm distance from the SRF origin. The other has 1.958 mm element pitch, in order to minimize the side-lobes for a source located at 308 mm. The characteristics of the two masks are presented in Table 1.

**Table 1.** Main characteristics of the two MURA Rank 19 masks

| Mask type | Number of elements | Diameter of elements (mm) | Surface (mm$^2$) | Element pitch (mm) | Ideal source distance (mm) |
|---|---|---|---|---|---|
| 19R-1821 | 37×37 | 1.7 | 67.23×67.23 | 1.821 | 160 |
| 19R-1958 | 37×37 | 1.7 | 72.19×72.19 | 1.958 | 308 |

Two cylindrical glass containers filled up to 9mm with $^{99m}$Tc radioactive uniform solution and a cylindrical gel solution of $^{57}$Co, 1.5 MBq activity each, acting as radio-traced sentinel node dummies, were arranged separately in several positions (X,Y,Z) within the system Fully Coded Field of View (FCFOV$_S$) [3]. The dimensions of the first source (type A) were 24mm in diameter and 9mm height, the dimensions of the second (type B) were 11mm diameter and 9mm height and the dimensions of the $^{57}$Co gel (type C) were 30mm diameter and 32mm height. Mask 19R-1821 was employed for source distances of 160 mm, 207 mm and 231 mm from the origin of the SRF and mask 19R-1958 was employed for a source distance of 308 mm. The distance between the centers of the PSD was set at 172mm. This setup provides a minimum operation height of 84 mm and a wide FCFOV$_S$ of about 117x117mm$^2$ at 160mm and 344x344mm$^2$ at 308mm. The counting time for each measurement was 60 sec.

**2.1 Resolution of the system for extended sources**

In order to study the resolution of our system, we summed the shadowgrams produced by placing the same source at two neighbor positions. In this subsection, experimental results for sources of type A and C are presented, because they are more extended and the effects due to this fact are more pronounced. For a given mask, when the sources are not at the optimum distance from the detectors plane, the resolution degrades and side lobes appear in the correlation matrix [8]. In addition, the presence of the side lobes of one source may coincide with the main peak of the other and in this way influences its Full Width at Half Maximum (FWHM) causing the degradation of the resolution. This can be seen in Figures 2b and 2c, as the volume and the lateral distance of the sources are changed in such a way that their apparent angle remains constant although the coordinate Z of the sources is different. When the two sources are placed at 308mm distance, the FWHM of the peaks of the correlation matrix decreases, so they become sharper, due to the smaller magnification of the mask element. This can be seen in Figure 2a compared with Figure 2b.

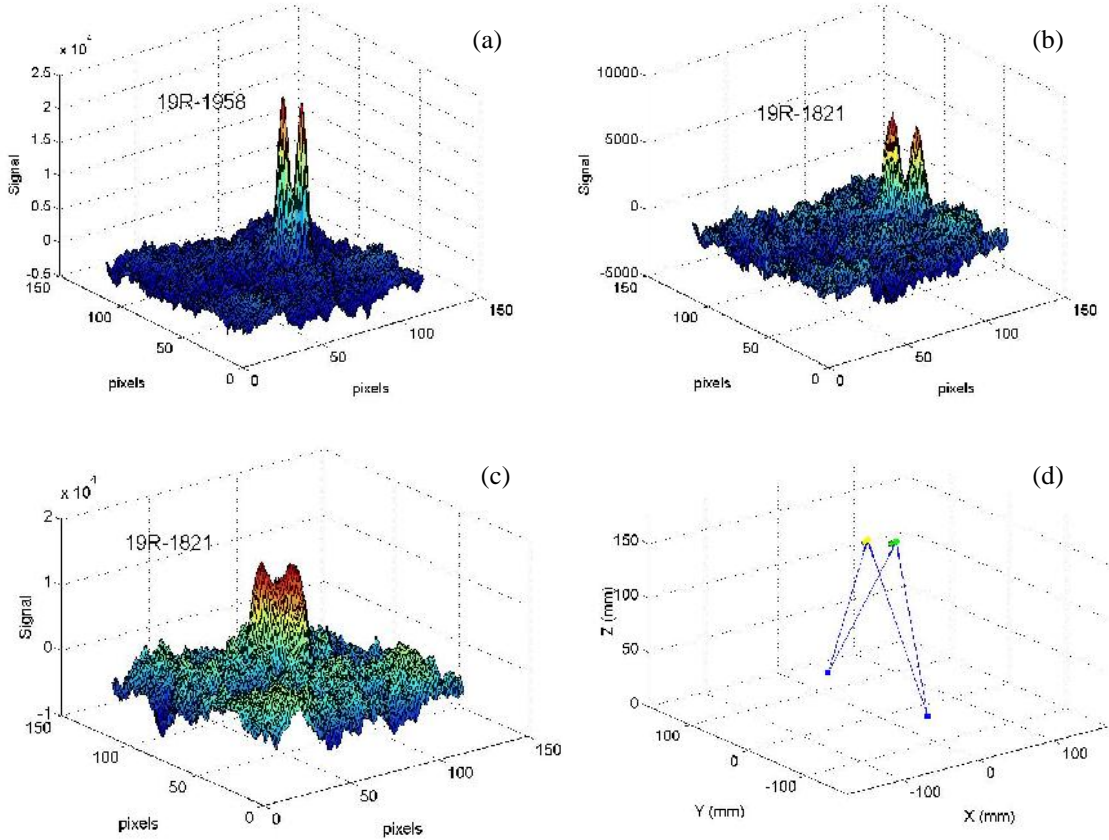

**Figure 2.** The correlation matrix for two cylindrical sources type A positioned at the center of FCFOV$_S$, (a) at Z=308mm and with distance between their centers 42 mm, (b) at Z=160mm and with distance between their centers 28mm. (c) The correlation matrix for two cylindrical sources type C positioned at Z=231mm and with distance between their centers 43mm. (d) The experimental 3D estimation of the centers of two sources of type A for the case described in (b) has an accuracy < 5%, after the triangulation. In each graph, the employed mask is indicated.

## 2.2 Accuracy of localization of extended sources

In order to measure the accuracy and the precision of the system in the presence of extended sources a series of experiments had been held. The sources of type A and B placed separately at two different Z, at several lateral distances $\sqrt{X_{real}^2 + Y_{real}^2}$ in the FCFOV$_S$. For each position, the estimation of its location was yield through triangulation and the source location accuracy |R|, presented in Figure 3, was calculated as:

$$|R| = \sqrt{(X_{real} - X_{estimated})^2 + (Y_{real} - Y_{estimated})^2 + (Z_{real} - Z_{estimated})^2}$$

where *real* and *estimated* are the actual and the estimated by triangulation coordinates of the center of the extended source, respectively, both with respect to SRF. Figure 3 shows that, for a given mask type, |R| deteriorates as the coordinate Z of the source is increased, from 6.1±0.61mm for Z=160mm, to 7.81±0.75mm for Z=220mm. In all cases the resulting |R| lays within the volume of the source and it is better than 5%. Simulations presented in Section 3.2 exhibit the same behavior (see Figures 6e and 6f).

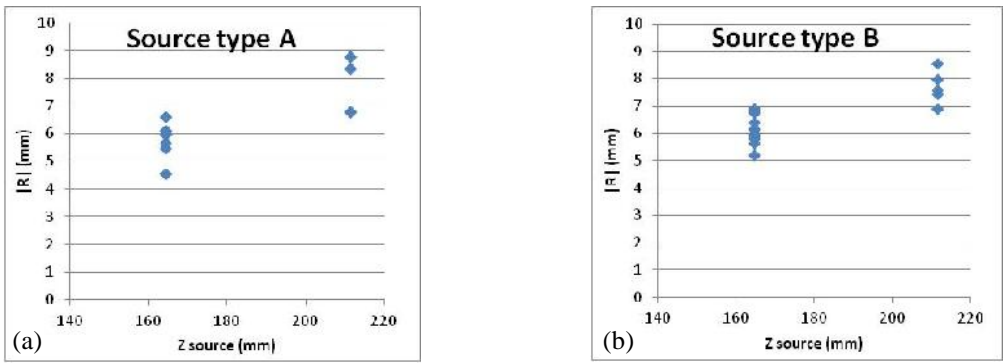

**Figure 3.** (a) |R| dependence on the Z coordinate of the type A source. (b) |R| dependence on the Z coordinate of the type B source . The same mask type 19R-1821 was used in both cases.

## 2.3 Signal to Noise Ratio throughout the FCFOV$_S$

The Signal-to-Noise ratio (SNR) is defined as the ratio of the net peak of the correlation matrix divided by the standard deviation of the noise in the correlation matrix. SNR deteriorates slightly when the source moves from the center towards the edge of the FCFOV (Figure 4), due to the deformed projection of the spherical elements on the shadowgram, as the source irradiates them from a non-perpendicular direction. For the data of Figure 4, the type A source located at Z = 308 mm and a single mask – PSD system, with the mask 19R-1958, was employed.

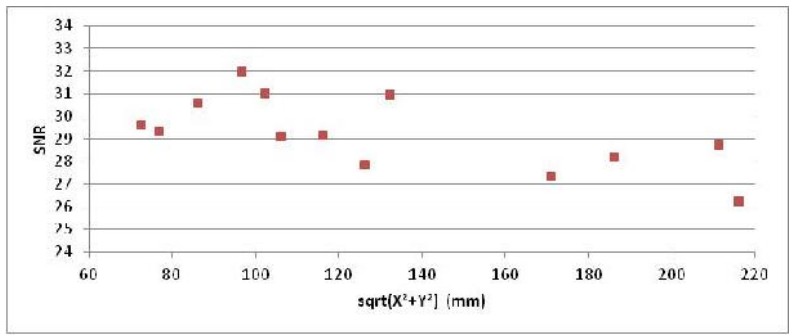

**Figure 4.** The SNR versus the lateral source distance from the center to the edge of the FCFOV for a single 19R-1958 mask – PSD system. The type A source is located at Z = 308mm.

## 3. Simulation Setup, Methods and Results

A fast simulation program [3, 4] has been developed to provide flexibility in changing the characteristics of the extended sources. Thus, we studied the dependence of the SNR and the FWHM of the correlation matrix peak on: (a) the dimensions of the source, (b) the distance Z of the source from the detectors plane, and (c) the distance of the source from the center to the edges of the FCFOV. The error bars shown in the plots of this section represent the standard deviation of the results of five-time repeated simulation experiments.

A single mask-PSD system was simulated, with the center of the PSD located at (X,Y,Z) = (0, 0, 0) mm and the mask placed at 20 mm above the PSD. The mask-PSD system was hit by randomly generated photons emanating from the volume of the extended source. Two kinds of 3-D extended sources were simulated, as N2 and N1 stages of lymph nodes cancer: the one with activity distributed uniformly within a cube with edge equal to d, the other with normally distributed activity with the 99.7% of the marginal distribution in each dimension confined to an interval with length equal to d. Consequently, the standard deviation ( ) for the uniform distribution is equal to $d/(2 \cdot \sqrt{3})$ mm and for the normal distribution is equal to $d/(2 \cdot 3)$ mm.

### 3.1 SNR and FWHM dependence on the dimensions of the extended source

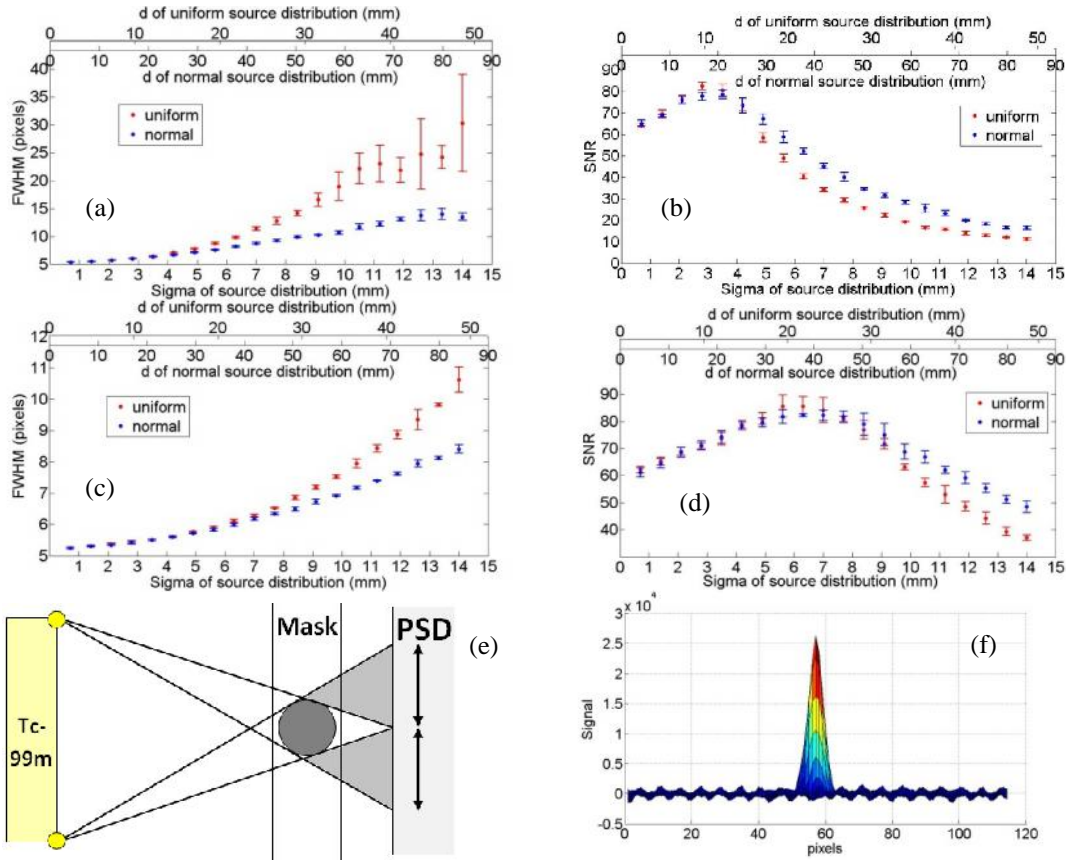

**Figure 5**: The FWHM (a) and the SNR (b) as a function of the   of the spatial distribution of the source for mask 19R-1821 and Z = 160mm. The FWHM (c) and the SNR (d) for mask 19R-1958 and Z = 308mm. In (e) the penumbra becomes larger as the dimension of the source increases and blurs the shade of a single element. In (f) the repeated structure of the noise on the y-z projection of the correlation matrix of a point-like source is shown. This structure is reduced as the condition shown in (e) is reached (size analogies are not real, for a more comprehensive illustration.)

The source was located at (X,Y,Z) = (0,0,160) mm, with mask 19R-1821 and at (X,Y,Z) = (0,0,308) mm with mask 19R-1958. The   of the spatial activity distributions was increased from 0.7 to 14 mm. The resulting FWHM expands slightly and its uncertainty rises, as the source becomes more extended (Figures 5a and 5c).The SNR reaches a maximum at    3mm of the activity distribution for Z source at 160 mm and at    6mm of the activity distribution for Z source at 308 mm (Figures 5b and 5d). A similar behavior appears in the dependence of the SNR on the distance Z of the source (Figures 6b and 6d). This is due to the superposition of the changing-magnified projection of the mask element and its penumbra. Specifically, the maximum SNR arises for d = 10 mm when the source is at Z = 160 mm and for d = 22 mm, when Z = 308 mm. For these conditions the shades of the mask elements shrink (Figure 5e) and the shadowgram is composed mostly by overlapping penumbras.  Therefore the repeated structure on the correlation matrix, shown in Figure 5f, vanishes.

**3.2 SNR, FWHM and |R| dependence on the distance of the source from the detectors**

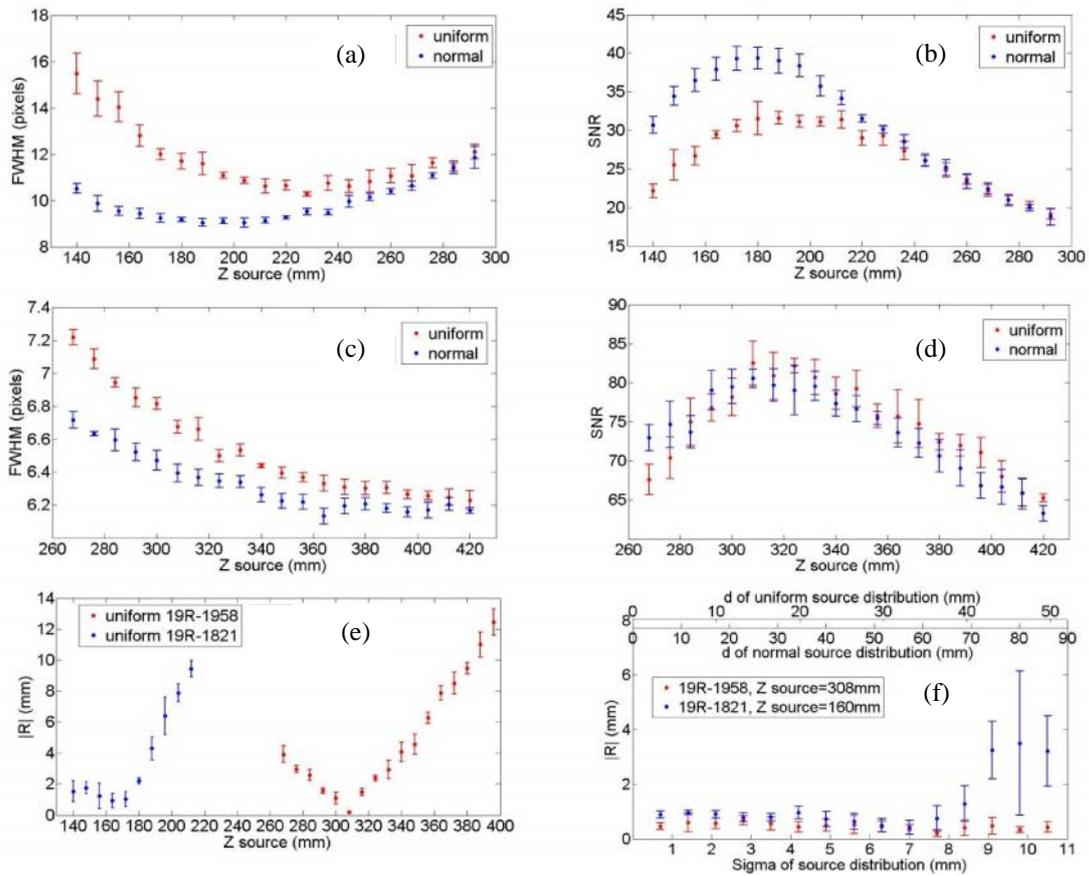

**Figure 6**: The FWHM (a) as a function of the source distance Z for mask type 19R-1821 and (b) the SNR. The FWHM (c) as a function of the source distance Z for mask type 19R-1958 and (d) the SNR. The |R| dependence on source distance Z for mask type 19R-1821 and mask type 19R-1958 (e). The |R| dependence on the   of the source with uniformly distributed activity for mask type 19R-1821 and mask type 19R-1958 (f).

A hot-spot with = 8mm, located at (X,Y) = (0,0) mm, irradiated a single mask-PSD system, while its distance Z from the detector plane was varied from 140 mm to 292 mm for mask type 19R-1821 and from 268 to 420mm for mask type19R-1958. The FWHM decreases as Z increases, due to the decrease of the magnification of the mask element (Figures 6a and 6c). For mask type 19R-1958 (Figure 6c) the change is slighter, due to the smaller span of element's magnification. For mask type 19R-1821 (Figure 6a) and when the Z rises above 230mm, an increase of FWHM is observed. As the extended source appears more point-like to the mask-PSD system when its distance from it increases, the value of the SNR for mask 19R-1958 is greater than the value of the SNR for mask 19R-1821 (Figure 6b and 6d). Let as note that we took care to adjust the activity of the source, so that the number of photons that reach the PSD remains constant, although the Z of the source changed.

In addition, a pair of mask-PSD systems, like the one in Figure 1, was simulated to estimate the 3D location of the hot spot via triangulation and thus study the deterioration of accuracy. Figure 6e shows that the |R| becomes minimum (maximum accuracy) for the ideal Z, i.e. that which results to a shadowgram with 19x19 element projections on it. In accordance with the experimental data (Figures 3a and 3b), |R| deteriorates for source distance higher than 160mm, when mask type 19R-1821 is used. The |R| dependence on is shown Figure 6f. For mask type 19R-1821, |R| deteriorates when the source dimension becomes larger than 35 mm. This happens because the peak on the correlation matrix becomes too wide and the fitting procedure cannot find accurately its center.

### 3.3 SNR and FWHM throughout the FCFOV$_S$

The FWHM and the SNR of the correlation matrix peak were studied as a function of the distance of the source from the center to the edges of the FCFOV$_S$. A hot-spot, with = 8mm, was located at Z = 160mm for mask type 19R-1821 and at Z = 308mm for mask type 19R-1958, while its lateral distance was varied from 0 to the edge of the FCFOV of one mask-PSD system. Also in this case, the activity of the source was adjusted so that the number of photons reached the mask is the same for both Z distances (i.e. 160 mm and 308 mm). The FWHM remains almost invariant, while its value is higher for the source with normally distributed activity than the one with activity distributed uniformly. The reason is that a source with normally distributed activity looks more point-like to the detector. In agreement with the experimental data presented in Figure 4, the SNR deteriorates slightly when the source lateral distance from the center of the FCFOV increases. In addition, the SNR value is higher for the source at Z=308mm, because the hot-spots appears more like a point-source to the detector.

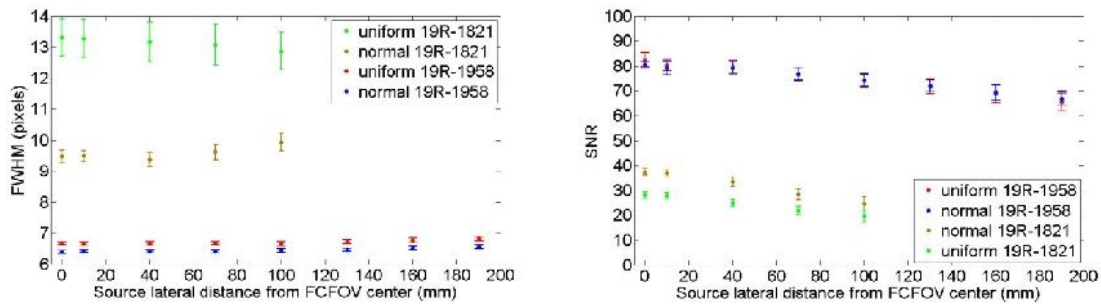

**Figure 7**: The FWHM (left) and the SNR (right) variations throughout the FCFOV (lateral distance) for a single mask-PSD system. A source with = 8mm, either with a uniform or normal distributed activity, is employed at the optimum distance for both mask types.

## 4. Conclusions

Our system is able to localize $^{99m}$Tc extended hot spots of up to a few centimeters wide in diameter within their volume dimensions and with an accuracy of better than 5%. For the utilized masks (MURA rank 19), the accuracy of localization is optimized when the magnification of the mask elements is such that a shadowgram of 19x19 elements appears on each detector. The resolution for two hot spots cylinders with diameter 24mm and height 9 mm is 28 mm when they are at a distance of Z=160 mm from the detectors plane and 42 mm when they are at a distance of Z=308 mm, while the $FCFOV_S$ remains wide at about 75° x75°.

The behavior of |R|, SNR and FWHM of the correlation matrix versus the topology of the hot spots was extensively studied with simulations. A notable result is that the SNR for an extended source under certain conditions is higher than that for a point source under the same conditions. In addition, the results of the aforementioned studies can be useful in the development of reconstruction methods of the 3D shape of hot spots.

## Acknowledgments

This research has been funded by NATO (SfP-984705) SENERA project and by the FP7-SEC-218000 COCAE project.